\documentclass[sigconf]{acmart}

\usepackage{balance}
\usepackage{booktabs} 

\begin{document}
\title{Mining and Forecasting Career Trajectories of Music Artists}

\copyrightyear{2018} 
\acmYear{2018} 
\setcopyright{acmcopyright}
\acmConference[HT '18]{29th ACM Conference on Hypertext and Social Media}{July 9--12, 2018}{Baltimore, MD, USA}
\acmBooktitle{HT '18: 29th ACM Conference on Hypertext and Social Media, July 9--12, 2018, Baltimore, MD, USA}
\acmPrice{15.00}
\acmDOI{10.1145/3209542.3209554}
\acmISBN{978-1-4503-5427-1/18/07}

\author{Shushan Arakelyan}
\affiliation{%
  \institution{USC Information Sciences Institute}
}
\email{shushan@isi.edu}

\author{Fred Morstatter}
\affiliation{%
  \institution{USC Information Sciences Institute}
}
\email{fredmors@isi.edu}

\author{Margaret Martin}
\affiliation{\institution{USC Information Sciences Institute}}
\email{mart586@usc.edu}

\author{Emilio Ferrara}
\affiliation{\institution{USC Information Sciences Institute}}
\email{ferrarae@isi.edu}

\author{Aram Galstyan}
\affiliation{\institution{USC Information Sciences Institute}}
\email{galstyan@isi.edu}

\renewcommand{\shortauthors}{S.Arakelyan et al.}

\begin{abstract}
Many musicians, from up-and-comers to established artists, rely heavily on performing live to promote and disseminate their music. To advertise live shows, artists often use concert discovery platforms that make it easier for their fans to track tour dates. In this paper, we ask whether digital traces of live performances generated on those platforms can be used to understand career trajectories  of artists. First, we present a new dataset we constructed by cross-referencing data from such platforms. We then demonstrate how this dataset can be used to mine and predict important career milestones for the musicians, such as signing by a major music label, or performing at a certain venue. Finally, we perform a temporal analysis of the bipartite artist-venue graph, and demonstrate that high centrality on this graph is correlated with success. 
\end{abstract}

%
%

\ccsdesc[500]{Information systems~Data mining}
\ccsdesc[300]{Information systems~Web mining}
\ccsdesc[500]{Networks~Online social networks}
\ccsdesc[500]{Human-centered computing~Collaborative and social computing}
\ccsdesc[300]{Computing methodologies~Machine learning approaches}
\ccsdesc[300]{Computing methodologies~Network science}

\keywords{networks, art and music, multidisciplinary topics and applications}

\maketitle
\section{Introduction}
Live performances are a crucial part of the life of a music artist. According to a recent industry report,\footnote{https://www.statista.com/statistics/491884/live-music-revenue-usa/} the revenues from live performances in the US have grown from \$8.72B in 2012 to \$9.94B in 2016, and are projected to reach almost  \$12B by 2022. A recent study discovered a connection between live events and increased digital listenership~\citep{ternovski2017social} (which is the second highest source of income for a band after live performances). In light of this, it becomes increasingly more important for artists to be able to understand what milestones matter to accomplish the dream of a professional career: playing at top venues goes hand-in-hand with getting more digital listeners, which in turn may increase their likelihood of being signed with major music labels. 

In this work, we aim to determine whether it is possible to model and predict these career trajectories under the emerging framework of \textit{Science of Success}~\cite{clauset2017data,fortunato2018science}: recent work studying how careers in different fields, as well as individual and team success, can be predicted early by leveraging records of performance from digital traces. This data-driven framework has been applied to domains as diverse as education and academia~\cite{kennedy2015predicting,sinatra2016quantifying,hosseinmardi2018discovering}, (e)sports~\cite{cintia2013engine,cintia2015network,zuber2015motivational,yucesoy2016untangling,sapienza2017performance}, social media~\cite{szabo2010predicting,bandari2012pulse,ma2013predicting,ferrara2014online}, culture~\cite{agreste2015analysis,yucesoy2018success}, and even the entertainment industry~\cite{park2016style,rossetti2017forecasting}. 

In light of these promising results, we pose the question: is it possible to find open data to understand and forecast careers and success in the music industry?
To accommodate the increasing demand of music artists to get their message out to their fans, specialized sites like \textit{Songkick} and \textit{Discogs} have sprung up to create centralized repositories of music events and music artists. These sites contain rich metadata about the artists themselves as well as the concerts they perform. They allow the artists to attract interests in their concerts. Indirectly, this goldmine also allows researchers to model the music industry dynamics.

\subsection*{Research Problem}
In this paper, we are interested in the problem of characterizing and understanding the career trajectories of the artists across different genres. Toward this goal, we analyze a large-scale longitudinal data of musical events occurring at various venues worldwide.  

Specifically, we  address the following research questions: 

\begin{enumerate}
\item Is the choice of venues where an artist performs correlated with the eventual success of that artist (for a given definition of success)? If so, can we leverage those correlations to forecast success? 
\item Can we predict which venues an artist/band will perform based on the history of his/her/their past performances? 
\item How do we measure the relative importance of performances in specific venues and their impact on career trajectories, and how do we jointly characterize \textit{influential} artists and venues? 
\end{enumerate}

\subsection*{Contributions of this Work}
Our main contributions are summarized as follows: 
\begin{itemize}
\item We construct and present a new dataset by collecting all of the artists and concerts from the {\em Songkick} platform, and supplement this dataset with information from {\em Discogs}, which contains more granular details about the artists---such as their discographies.\footnote{The dataset is available at \url{https://github.com/shushanarakelyan/forecasting_success}
}

\item We define a measure of success based on whether an artist has signed a contract with one of the major music record labels, and propose a forecasting task to differentiate between career trajectories of artist based on this measure of success. 

\item We demonstrate the viability of forecasting future performances of artists, and therefore their success, based on the history of past performances.

\item We propose a centrality measure suited for the bipartite artist-venue network and demonstrate that it correlates strongly with the venue reputation. 

\end{itemize}

The rest of the paper is organized as follows. After describing related work in Section~\ref{sec:related}, we describe the dataset in Section~\ref{sec:dataset} and provide its basic statistics in Section~\ref{sec:statistics}. In Section~\ref{sec:analysis} we define three related tasks - forecasting artist success, predicting future events by artist at specific venues, and identifying influential artists and venues -  describe our approach for addressing those tasks, and present results. We conclude the paper by summarizing our main findings in Section~\ref{sec:conclusion}. 

\section{Related Work}
\label{sec:related}

Quantifying and forecasting success refers to the broader body of work that attempts to discover the patterns and performance trajectories that correlate with certain desirable outcomes: from forecasting highly-cited academic authors and papers \cite{wang2013quantifying,ke2015defining} to predicting future Nobel Prize winners \cite{mazloumian2011citation}, from uncovering successful fund-raising campaigns \cite{mitra2014language}, to early identifying the next top model \cite{park2016style}, or movie box office hit \cite{dellarocas2007exploring}. The new field of \textit{Science of Success} brings a strong data-driven perspective on applied forecasting problems set in the real world. 

\citet{judge1999big} postulated that career success has intrinsic cues, like the person's own perception of success and self-satisfaction, and extrinsic ones, like awards, recognition or achievements. Since judgments about success in a creative profession like music are unavoidably subjective, we don't consider intrinsic factors and focus on objectively observable career accomplishments only. 

Music industry criteria called ``traditional markers of artist success''~\cite{evans2013constitutes}, like performance opportunities, labels, charts, awards, sales of recorded music or airplay, provide us with a number of possible directions for defining success of music artists. However, digitization has shaken these traditional markers---digital music has been linked to fall in record sales, airplay and charts no longer adequately measure popularity, given numerous streaming services and listenership outside of them---views on YouTube and/or illegal file-sharing. Given this, some researchers look at the popularity of music artists on digital delivery platforms like Last.fm, and formulate a forecasting problem to predict new song hits from the early adoption patterns of music listeners~\cite{rossetti2017forecasting}.

Success in post-digital music world can still be adequately represented by contracts with major  labels. Music record labels are still important players in the industry---even though theoretically digital technologies allow artists to perform production, promotion and sales on their own, practically this doesn't happen very often~\cite{mclean2010myths}. Hence, in this work, forecasting success is operationalized as predicting the artists that are going to be signed by a major music recording label. To the best of our knowledge, this is a novel formulation that has not been presented in the literature before.

From a methodological perspective, our work is rooted on a blend of machine learning and network science techniques. 
We focus in particular on a broad class of problems often referred to as \textit{link mining} (a.k.a. \textit{link prediction}). 
Link mining is the problem of discovering new (unforeseen) edges in a graph. 
Typical possible applications are either network reconstruction ~\cite{clauset2008hierarchical,guimera2009missing}, or modeling the evolution of a network ~\cite{kashima2006parameterized,zhu2016scalable,bliss2014evolutionary}. One common operationalization of link prediction is finding pairs of nodes that have high probability of being connected. This often translates into measuring node similarities, as mentioned by Liben-Nowell and Kleinberg ~\cite{liben2007link}. However, other authors \cite{kunegis2010link} noted that using traditional link prediction on bipartite graphs is not straightforward and often produces counterintuitive results. In order to address this shortcoming, some authors proposed modified similarity metrics \cite{liben2007link,kunegis2010link}, or used techniques from recommender systems, such as low-rank matrix factorization and collaborative filtering \cite{acar2009link,buza2013application}, and supervised learning approaches \cite{pavlov2007finding,benchettara2010supervised}. We follow the example of those authors and use collaborative filtering and recommender systems inspired methods to perform link prediction for our task. 
In the results section, we will show how to leverage \textit{BiRank}~\cite{he2017birank}---a modification to the \textit{PageRank}~\cite{page1999pagerank} algorithm that tunes it towards bipartite graphs---to measure and predict the popularity of the artists and venues. 

\section{Dataset}
\label{sec:dataset}

\textbf{\textsc{Songkick}}\footnote{\url{https://www.songkick.com/}} is a concert-discovery platform that aims to link fans to artists' events. It contains information about over 6 million concerts (and other music events like festivals), the artist(s) that perform at each event, and the venue where each event takes place. The ``gigography'' of an artist is the term that Songkick uses to refer to all of that artist's events. 

Songkick data can be accessed through their website or via their API, which allows querying any artist's gigography. 
Songkick is our main repository of information for music events.

\textbf{\textsc{Discogs}}\footnote{\url{https://www.discogs.com/}} is a music database that contains cross-referenced discographies of artists and labels. Each recording, artist, or label in Discogs can be uniquely identified by their IDs. Discogs provides separate data dumps\footnote{\url{https://data.discogs.com/}} for artists, labels, and recordings. We used recordings data dump from May 1, 2017 to obtain artist and label IDs associated with each release. This data dump contains more than 8 million recordings. Most of the recordings have information about their release dates, and thus allow tracking the history of releases with different labels for each artist. 



\subsection{Data Collection}
Songkick does not provide a lookup directory of artists, nor there is a direct mechanism to get all gigographies. For getting Songkick artist IDs we queried artist names present in Discogs' recordings data dump. As a result, all of the artists in our dataset have at least one recording on Discogs. This can be either self-recorded or recorded under a contract with a music label. This strategy avoids introduction of bias towards artists that did not publish any recordings, which are therefore excluded from our analysis.

The Songkick API call returns a list of possibly relevant artists, allowing for some inexact name matching. 
We processed the API output to retain data on artists that exactly matched the Discogs artist name.

From this name match we obtained artist IDs, and used them for another round of API calls, to get the gigographies of each artist. For each concert in the gigography, we extracted the following information: ID, date, city, country, state (if applicable), latitude and longitude of the venue, venue ID and venue name, name of the event and its popularity score as calculated by Songkick. 

For every event there is information about billing for each artist, i.e., whether that artist was a headliner or a support artist at the concert. However, we did not consider headliners and support artists separately in the analysis presented further.

Collected data was organized into separate artist, event, and venue data frames. Each artist is indexed by its Songkick and Discogs IDs. Venues and events are indexed by their Songkick IDs. There are also several lists of cross-references: mapping venues to the events that happened there, and events to the venues where they took place. A similar mapping is available for events and artists, and releases and artists.


\subsection{Data Preprocessing}
Due to the fact that the goal of Songkick is connecting fans to their favorite artists through concerts, the platform puts less relevance on events that occurred prior to their inception. Songkick was founded in 2007 and there is a noticeable increase in the number of artists that have their earliest concerts recorded on Songkick in 2007 or later (see Figure~\ref{fig:artist_earliest_performance_year} in the next section). For the sake of data completeness, we focus only on artists that have their first record of performance in 2007 or later. By doing so, we aim to retain only the artists who used Songkick to inform their fans about upcoming events, thus avoiding the use of possibly incorrect backdated data.


In this paper we consider an artist that has one or more recordings with one of the major labels (a.k.a., ``Big Three''/Four/Five/Six),\footnote{ \url{https://en.wikipedia.org/wiki/Record_label\#Major_labels}} or their subsidiaries, ``successful'', we provide a more detailed explanation for this choice in Section~\ref{sec:forecasting_artist_success_intro}. Conveniently, each music record label in Discogs has information about its sub-labels and its parent label, if such exist. This allowed tracking all subsidiaries of the major labels. We assume the first time an artist releases a recording with such a label to be the change point in their career. We are interested in researching the trajectory of artists before the change point and ideally being able to forecast the change point.

Finally, we wanted to make sure that we have enough data about successful artists in the early stage of their career. Thus, in the last preprocessing step, we removed every artist and venue that has less than 10 concerts associated with them before the change point. This also takes care of venues that may have been used for occasional events, or artists with short-lived careers.  

\section{Statistics}
\label{sec:statistics}

In the following we provide some statistical analysis of our dataset. 
The dataset contains 645,507 concerts, 13,912 artists, and 11,428 venues, collected for the time frame between 2007 and 2017. Artists in the dataset are associated with 39,641 distinct record labels, 286 of which are major labels, or their subsidiaries. One condition to be labeled as a ``successful'' artist in our study is to have recorded at least one album under any of these 286 recording labels.

\begin{figure}[!t]
	\centering
		\includegraphics[width=0.5\textwidth]{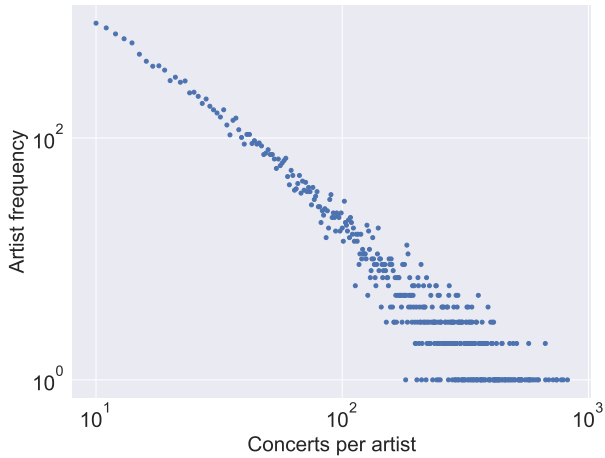}\\
        \includegraphics[width=0.5\textwidth]{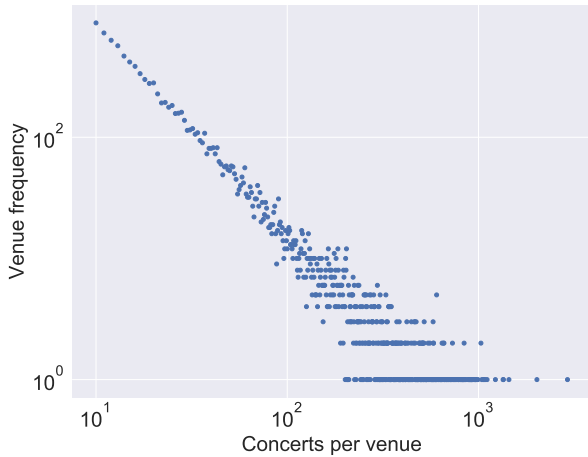}
    \caption{Heavy-tailed distributions of the number of concerts per artist (upper panel) and per venue (lower panel).}
	\label{fig:concerts_per_artist_and_venue}
\end{figure}

Figure~\ref{fig:concerts_per_artist_and_venue} depicts distributions of the number of concerts per artist and number of concerts per venue. Both distributions are very broad and heavy tailed, with few active artists and venues hosting many events, and a very large set of artists and venues associated with very few events. 

\begin{figure}[!h]
	\centering
		\includegraphics[width=0.5\textwidth]{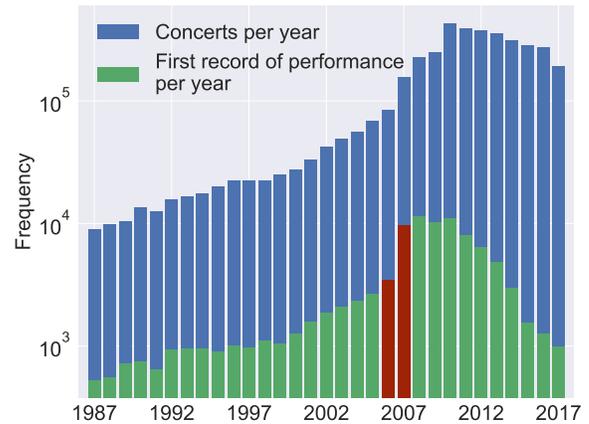}
    \caption{Total number of concerts per year in the dataset and number of artists that first appear on Songkick in a given year. The red bars illustrate the sudden change from 2006 to 2007 in the number of artists that first appeared on Songkick during those years. This can be explained by the fact that Songkick was founded in 2007. Data before 1987 are very limited thus not included in this illustration. \\   
    }
	\label{fig:artist_earliest_performance_year}
\end{figure}

In Figure~\ref{fig:artist_earliest_performance_year} we show the dynamics of the number of events and number of artists from 1987 to 2017. As already mentioned, there is a significant increase in the number of artists that have their earliest concerts recorded on Songkick in 2007 or later. From Figure~\ref{fig:artist_earliest_performance_year} it can be seen that the total number of concerts per year peaked in 2010. 

\begin{figure}[!h]
	\centering
     \includegraphics[width=0.5\textwidth]{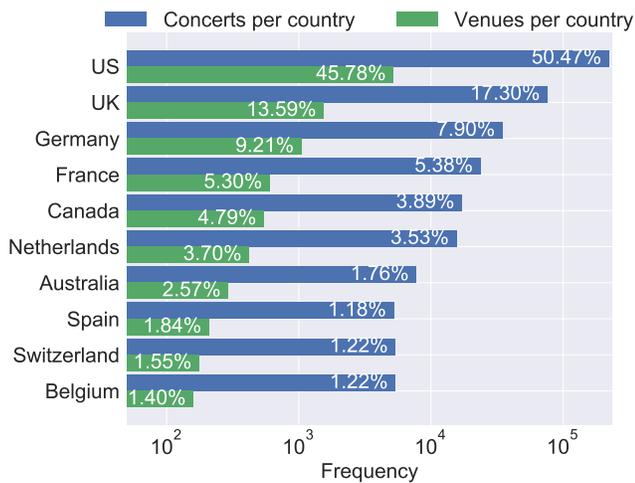}
    \caption{Log-scale distribution of concert frequencies in (i) the top 10 most active countries, and (ii) the number of distinct venues in those countries. A disproportionate preference toward English-speaking countries can be observed in the Songkick data, with United States, United Kingdom, and Australia cumulatively accounting for nearly 70\% of the total events, and over 60\% of the total venues. \\
    }
	\label{fig:geographic_distributions_of_concerts}
\end{figure}

Next, we look at the geographic distribution of venues in the dataset. There are 63 different countries with at least one event, which, for the most part, are in North America and Europe. Almost half of all venues are located in the United States, where also more than half of all concerts happened. The second highest in both number of concerts and number of venues is the UK. Figure~\ref{fig:geographic_distributions_of_concerts} demonstrates distribution of venues and concerts in the 10 most frequently occurring countries.

\begin{table}[!t]
\centering
\caption {Some of the most frequent n-grams extracted from sequences of artists' performances. Double-sided arrows indicate that these routes are frequently found in the data in both directions.}
\captionsetup{width=\textwidth}
\begin{tabular}{p{.46\textwidth}}
\toprule
\textbf{Frequent routes that artists follow}\\
\midrule
$\textnormal{San Diego} \leftrightarrow \textnormal{Los Angeles} \leftrightarrow \textnormal{SF Bay Area} \leftrightarrow \textnormal{Portland} \leftrightarrow \textnormal{Seattle}$\\
\midrule
$\textnormal{Portland} \leftrightarrow \textnormal{Seattle} \leftrightarrow \textnormal{Boise} \leftrightarrow \textnormal{Salt Lake City} \leftrightarrow \textnormal{Denver}$\\
\midrule
$\textnormal{Chicago} \leftrightarrow \textnormal{Toronto} \leftrightarrow \textnormal{Montreal} \leftrightarrow \textnormal{Boston/Cambridge} \leftrightarrow \textnormal{New York}$\\
\midrule
$\textnormal{Washington} \leftrightarrow \textnormal{Philadelphia} \leftrightarrow \textnormal{New York} \leftrightarrow \textnormal{Boston/Cambridge}$\\
\midrule
$\textnormal{London} \leftrightarrow \textnormal{Birmingham} \leftrightarrow \textnormal {Manchester} \leftrightarrow \textnormal{Glasgow}$\\
\midrule
$\textnormal{Brisbane} \leftrightarrow \textnormal{Sydney} \leftrightarrow \textnormal {Melbourne} \leftrightarrow \textnormal{Adelaide}$\\
\midrule
$\textnormal{Austin} \leftrightarrow \textnormal{Houston} \leftrightarrow \textnormal{New Orleans} \leftrightarrow \textnormal{Atlanta}$\\
\bottomrule
\end{tabular}
\label{tab:frequent_routes}
\end{table}

If we look at more granular information about geolocation of artists' performances we can get an insight on actual spatial trajectories of artists. Particularly, we can look for frequent subsequences among the sequences of performances of all artists. As displayed in Table~\ref{tab:frequent_routes}, n-grams of length 4 and 5 show some frequent routes of cities that artists take while touring. Following the distribution of the venues and concerts in the dataset, most frequent routes mostly include US cities. 
As demonstrated in Table~\ref{tab:frequent_routes}, frequent routes contain clear patterns of artists performing in big cities on their way, while travelling from North to South or from East to West, etc.  

\section{Analysis and Results}
\label{sec:analysis}
To better illustrate the idea that the music artist career trajectory can be predicted from artist-venue interactions we formulated the following 3 tasks, discussed next:

\begin{itemize}
\item Task 1: Forecasting artist success;
\item Task 2: Event prediction;
\item Task 3: Joint discovery of influential artists and venues.
\end{itemize}
In the next subsections, we describe each of those tasks in more details, elaborate on our approach for addressing them, and present our results. 

\subsection{Task 1: Forecasting Artist Success}
\label{sec:forecasting_artist_success_intro}

Due to the nature of the partnership between artists and record companies, the bigger the recording label the more resources and opportunities it has to offer for its artists. Artists, nurtured by labels, have the chance to develop their sound, their craft, and their careers. Besides, record companies facilitate introductions to world-class producers, writers, and other performers, which can determine careers and bring huge rewards.

The recording industry has been marked by concentration and centralization for a while now. During the phase of consolidation in 1970s, most of the major labels were acquired by very few umbrella corporations or music groups. The Beatles, Frank Sinatra, Pink Floyd and even Maria Callas found prominence through those major record labels.
From 1988 till 2012 the number of  major record companies has decreased from six to three, as some of them got absorbed by the others. The remaining three major music groups, or the \textit{Big Three} (Sony BMG, Universal Music Group, and Warner Music Group), have held a large share of the world music production since 2012.

Because of the influence and patronizing that the major labels provide, we consider artists that have a recording with either the parent major label, or one of its subsidiaries, as \textit{successful}. We set to see if the rise to success can be predicted from a sequence of performances. Our goal in this task is, therefore, to identify successful artists from their career trajectories.

Ideally, we want to be able to identify such artists in a post-hoc manner. In other words, we want to detect the change that will lead to a release with a major label before the release itself happens. In the following discussion we refer to this task as \textit{forecasting}.

We also consider the simpler task of discriminating artists that are already successful in our setup from the ones that are not. We refer to this task as \textit{prediction}. 

\subsubsection{Experimental Setting}
For both forecasting and prediction tasks we used the \textit{affiliation matrix} of artists and venues. In such an affiliation matrix an artist is represented as a bag-of-words vector over the venues where the artist has performed. The entries in the matrix are the numbers of times the artists performed at the venue.
We used those vectors as features for the prediction and forecasting tasks.

In the forecasting task for any artist we did not include any concert that happened after the artist released their first recording with a major music label. However, for the prediction task we included those performances too. 

The classification labels (successful or not) were obtained by iterating over all the music labels that each artist has ever recorded with (this information was obtained from Discogs). If among these music labels there are either major ones or one of their subsidiaries, we assume that the artist was successful and label it as a positive instance---negative otherwise.

As a result of the procedure above, we labeled about 500 artists as successful, which is  3.6\% of the total number. It is worth noting that our labeling procedure yields a highly unbalanced dataset where the positive instances (successful artists) are very infrequent: this is in line with the commonsense notion of popularity in the music industry, where musicians that thrive with a professional career are exceptionally rare. 

\subsubsection{Metrics}
A natural choice for evaluating a success forecasting or prediction task is classification accuracy. However, due to high imbalance in the data, we need metrics that are more sensitive and account for under-represented classes. Such metrics are Precision, Recall and F1 score, as well as ROC AUC score, which we used for evaluation.

\subsubsection{Learning Models and Configuration}
For Task 1, we defined three simple models described next, and used them to carry out the forecasting and predictions exercises.

\textbf{Baseline}: We can intuitively connect success of the artist to the number of their performances. We picked a baseline that would prove or disprove this scenario by using the number of concerts, scaled by the maximum number of concerts by an artist, as a proxy for probability for becoming successful.

\textbf{Logistic Regression}: As a base classifier in both prediction and forecasting experiments we used Logistic Regression from the scikit-learn library~\cite{scikit-learn}. We used $L_2$ norm for regularization, and tuned one parameter, i.e., the inverse of regularization strength $C$.

\textbf{SVD}: Since the affiliation matrix we use has over 99\% sparsity (percentage of zero entries), dimensionality reduction techniques could yield prediction performance improvements by transforming sparse data into dense. We performed dimensionality reduction using Singular Value Decomposition (SVD). Via cross-validation we discovered that best results are achieved when we use 750 components in prediction task and 1000 components in forecasting task. 

For each model, we performed hyperparameter tuning via grid search with 3-fold cross validation on the training set. The results reported are obtained by using cross-validated average over 3 different train-test splits in 80-20 ratio. 

\subsubsection{Task Summary}
The results for this task are presented in Table~\ref{tab:success_prediction}. Suggested baseline shows existing correlation between the number of concerts and prediction label, and this correlation is stronger in prediction task than in forecasting task. Next, simple logistic regression achieves 0.22 F1 score on the forecasting task and 0.4 on the prediction task. We can see that while reducing dimensions increase ROC AUC and F1 scores by several points in forecasting task, its improvement for prediction task is marginal.

The improvement in performance on the prediction task indicates there is a difference in distributions of artist performances before and after they record their first album with a major music label. This suggests the existence of change points in careers that are caused by recording with major labels, which corroborates our notion of artist's  success. We expect that employing more sophisticated models for discovering change points would give better forecasting results.

\begin{table}
\centering
\caption {Precision (P), Recall (R), F1-score and  AUC for artist success forecasting (FCST) and prediction (PRED) tasks. We show results of logistic regression on full data (FCST/PRED LR) and with reduced dimensions (FCST/PRED LR+SVD)}
\captionsetup{width=\textwidth}
\begin{tabular}{p{.04\textwidth}p{.15\textwidth}p{.04\textwidth}p{.04\textwidth}p{.04\textwidth}p{.04\textwidth}}
\toprule
\textbf{Task} & \textbf{Model} & \textbf{P} & \textbf{R} & \textbf{F1} & \textbf{AUC} \\
\midrule
FCST & Baseline & 0.07 & 0.26 & 0.11 & 0.60 \\
FCST & LR & 0.18 & 0.29 & 0.22 & 0.74 \\ 
FCST & LR+SVD & 0.18 & 0.35 & 0.23 & 0.78 \\
\midrule
PRED & Baseline & 0.25 & 0.35 & 0.29  & 0.82 \\
PRED & LR & 0.36 & 0.45 & 0.40 & 0.86 \\
PRED & LR+SVD & 0.39 & 0.40 & 0.40 & 0.87 \\
\bottomrule
\end{tabular}
\label{tab:success_prediction}
\end{table}

\subsection{Task 2: Event Prediction}
Besides artist career trajectories, we are also interested in the overall dynamics of the network, where both venues and artists evolve and their influence changes as a result of constant interactions between venues and artists.  

To see if we can explain part of those interactions, we formulate the artist-venue link prediction task. As in the forecasting artist success task, we consider here two configurations---\textit{forecasting} and \textit{prediction}. For this task we used the same affiliation network as in the previous task, but since we are interested in predicting new or hidden edges, we only use a binary affiliation matrix here.

In the previous task prediction experiments were performed to test whether or not our suggested definition of success is viable. For $(artist-venue)$ link mining task, however,  we exercise prediction alongside to the forecasting to test for possible major temporal shifts in artists' behavior.

\subsubsection{Experimental Setting}
In the forecasting task, we looked for new $(artist,venue)$ links, or edges, based on the history of old ones. In particular, we used all performances from 2007 to 2015 as ``history'' (i.e., training data), and the performances in 2016 and 2017 as ``future'' (i.e., test set). We then went on and recursively removed all artists and venues that have less than 5 concerts associated with them in the training set. As a result we had 12,871 artists, 10,269 venues, 385,845 events in the training set and 43,122 events in the test set.

In the prediction task we kept the same set of artists and venues as described above for the forecasting task. We then randomly picked 20\% of all links and hid them in the test data, using the remaining 80\% for training purposes, similarly to a link prediction problem. Results reported are averaged over such three random splits. We binarized all the links as we are only interested in predicting new links, i.e. new venues, where artist performs.

\subsubsection{Metrics}
We measured the performance on this task using Area Under the Receiver Operating Characteristic curve (ROC AUC). 
One of the main advantages of this metric is the fact that it operates on rankings and is calculated for a range of thresholds, rather than prediction classes. This allows us to interchangeably use simple recommender system objectives for venue prediction.

\subsubsection{Learning Models and Configuration}
For Task 2, we decided to adopt some popular heuristic scores for link prediction, a simple matrix factorization technique and node similarity based model, all described in the following.

\textbf{Heuristic scores:}
Likelihood of a link existing between a pair of nodes is often approximated in terms of the number of common direct neighbors of that pair. However, a score calculated in this way will always be zero in a bipartite graph. Hence, we extended popular methods---Common Neighbors and Jaccard's coefficient---to use 2-hop neighbor sets of the pair instead of direct neighbors, as shown in Table~\ref{tab:heuristics}, where $\mathcal{N}(u)$ is defined as the set of direct neighbors of node $u$, and $\hat{\mathcal{N}}(u) = \cup_{v \in \mathcal{N}(u)} \mathcal{N}(v)$ is the set of \textit{neighbors of neighbors} of $u$. Another popular link prediction heuristic is Preferential attachment, which can be applied to a bipartite graph without any modifications. 

\renewcommand{\arraystretch}{1.7}
\begin{table}
\centering
\caption {Heuristic scores for link prediction in bipartite graph for node pair $(u, v)$. $\mathcal{N}(u)$ indicates direct neighbors of node $u$, $\hat{\mathcal{N}}(u)$ indicates neighbors of neighbors of $u$.}
\captionsetup{width=\textwidth}

\begin{tabular}{@{}c|c@{}}
\toprule
Common Neighbors ($CN(u, v)$) & 
$|
(\hat{\mathcal{N}}(u) \cap \mathcal{N}(v) ) \cup (\hat{\mathcal{N}}(v) \cap \mathcal{N}(u) ) 
|$\\
Jaccard's Coefficient & 
$\frac{
CN(u, v)}
{|
\hat{\mathcal{N}}(u) \cup \mathcal{N}(v) 
\cup 
\hat{\mathcal{N}}(v) \cup \mathcal{N}(u) |
}$\\
Preferential Attachment&
$|
\mathcal{N}(u)| \cdot |\mathcal{N}(v)
|$
\\
\bottomrule
\end{tabular}
\label{tab:heuristics}
\end{table}
\renewcommand{\arraystretch}{1}

\textbf{Matrix factorization:}
Link mining in a bipartite graph can be naturally presented as a recommendation task. For each artist we have a list of ``relevant'' venues---the ones where the artist performed. Using methods for collaborative filtering we can find latent features or representations of venues that make them relevant for certain artists. Based on these hidden representations, we can then predict which venues are most relevant for the artist. 

In this task, we used a simple yet popular collaborative filtering method based on matrix factorization---Singular Value Decomposition (SVD). To find the number of components for SVD, we used grid search---from 10 to 2000---and reported the result for 25.

\textbf{Node similarity:}
Building and using graph representations is another approach that is often employed for link prediction. In our experiments we leveraged Deepwalk \cite{perozzi2014deepwalk} for obtaining node representations and then used cosine similarity of a pair of nodes as an estimate for the probability of a link existing between them.\footnote{https://github.com/phanein/deepwalk}

Deepwalk is similar to training a Word2Vec model on a random walk sampled starting from every node in the graph. In our graph we gave preference to a larger number of short walks so we searched for the optimal number of walks of length 10. We report results for using 40 random walks. We then used cosine similarity of node representations as a proxy for probability of creating a new edge between those nodes. 

Hyperparameters like number of hidden components in SVD and Deepwalk parameters in this task were only tuned for prediction task. We then used the same values for forecasting task. All parameters were estimated via grid search with 5-fold cross-validation, with 20\% of all edges in each fold.

\subsubsection{Task Summary}
The results for the venue prediction task are presented in Table~\ref{tab:predict_venues}. As it can be seen, every method performs better on the prediction task than on forecasting, though for heuristic methods the improvement in performance is marginal. This hints that there might be a shift in artists' preferences for choosing a venue over time. It also indicates that while coarse statistics like Common Neighbors or Jaccard's coefficient are not affected much by those shifts, slightly more sensitive methods like SVD and node similarity, that rely on the inner structure of the graph, are affected more. Yet, either that structure is not expressive, or the methods are not powerful enough, neither of those methods performs better than heuristic scores. Interestingly, four models out of five give performance of around 0.9 ROC AUC on prediction task. Out of all the methods we tried, Preferential Attachment has the lowest performance for both tasks. 


\begin{table}
\centering
\caption {Results for $(artist, venue)$ link prediction task, measured in Area Under Receiver Operating Characteristics curve (AUC).}
\captionsetup{width=\textwidth}
\begin{tabular}{p{.07\textwidth}p{.27\textwidth}p{.07\textwidth}}
\toprule
\textbf{Task} & \textbf{Model} & \textbf{AUC} \\
\midrule
FCST & Common Neighbors & 0.87 \\
FCST & Jaccard's coef & 0.89 \\
FCST & Preferential Attachment & 0.79 \\
FCST & SVD & 0.81 \\
FCST & Node similarity & 0.84 \\
\midrule
PRED & Common Neighbors & 0.91 \\
PRED & Jaccard's coef & 0.90 \\
PRED & Preferential Attachment & 0.84 \\
PRED & SVD & 0.91 \\
PRED & Node similarity & 0.90\\
\bottomrule
\end{tabular}
\label{tab:predict_venues}
\end{table}

\begin{table*}[!h]
	\centering
	\caption{The most influential nodes of each class identified by BiRank.}
    \begin{tabular}{l c c}
    \toprule
    Rank & Artists & Venues \\
    \midrule
    1 & Frank Turner & The Observatory, Los Angeles, CA \\
    2 & Every Time I Die & The Masquerade, Atlanta, GA \\
    3 & Against Me! & The Bowery Ballroom, New York, NY \\
    4 & Reel Big Fish & Webster Hall, New York, NY \\
    5 & All Time Low & 9:30 Club, Washington, DC \\
    6 & The Black Dahlia Murder & House of Blues, Boston / Cambridge, MA \\
    7 & Hatebreed & Theater of the Living Arts, Philadelphia, PA \\
    8 & Future Islands & The Middle East Downstairs, Boston / Cambridge, MA \\
    9 & Halestorm & Vienna Arena (Arena Wien), Vienna \\
    10 & Hawthorne Heights & Brudenell Social Club, Leeds \\
    \bottomrule
    \end{tabular}
    \label{tab:influencers}
\end{table*}

\subsection{Task 3: Joint Discovery of Influential Artists and Venues}
In the previous tasks, we have attempted to classify an artist as about to be signed or not about to be signed. In this task we will investigate whether we can identify top artists and venues automatically by mining their performances. 

To measure the popularity of the artists and venues, we leverage BiRank~\cite{he2017birank}. This algorithm is a modification to the PageRank~\cite{page1999pagerank} algorithm that tunes it towards bipartite graphs. The algorithm iteratively identifies influential venues by observing which influential artists play at them. Simultaneously, it measures influential artists by measuring their frequency of playing at influential venues. 

Before running this algorithm, we set the initial ranking based upon the following measure:
\begin{equation}
	g_i = \frac{\log(N_i + 1)}{\sum_{a \in \mathcal{A}} \log(N_a + 1)},
\end{equation}
where $N_i$ measures the number of links to the node  $i$, $\mathcal{A}$ is the set of artists in the dataset, and $i \in \mathcal{A}$. This constitutes the artist's initial score. Similarly, we compute: 
\begin{equation}
	g_j = \frac{\log(N_j + 1)}{\sum_{v \in \mathcal{V}} \log(N_v + 1)},
\end{equation}
where $\mathcal{V}$ is the set of venues and $j \in \mathcal{V}$. With this initial seed score, we proceed to run the BiRank algorithm to identify the most influential nodes in each set. Finally, it is important to note that there is a temporal weighting in the links. Each link in the adjacency matrix has a weight of $\delta^{2017 - y_0}$, where delta is the decay parameter (set to 0.85 in the experiments), and $y_0$ is the year of the first link. We subtract this number from 2017 as this is the most recent year in the dataset. This experimental setup closely resembles that of~\cite{he2017birank}.

\begin{figure}
	\includegraphics[width=0.5\textwidth]{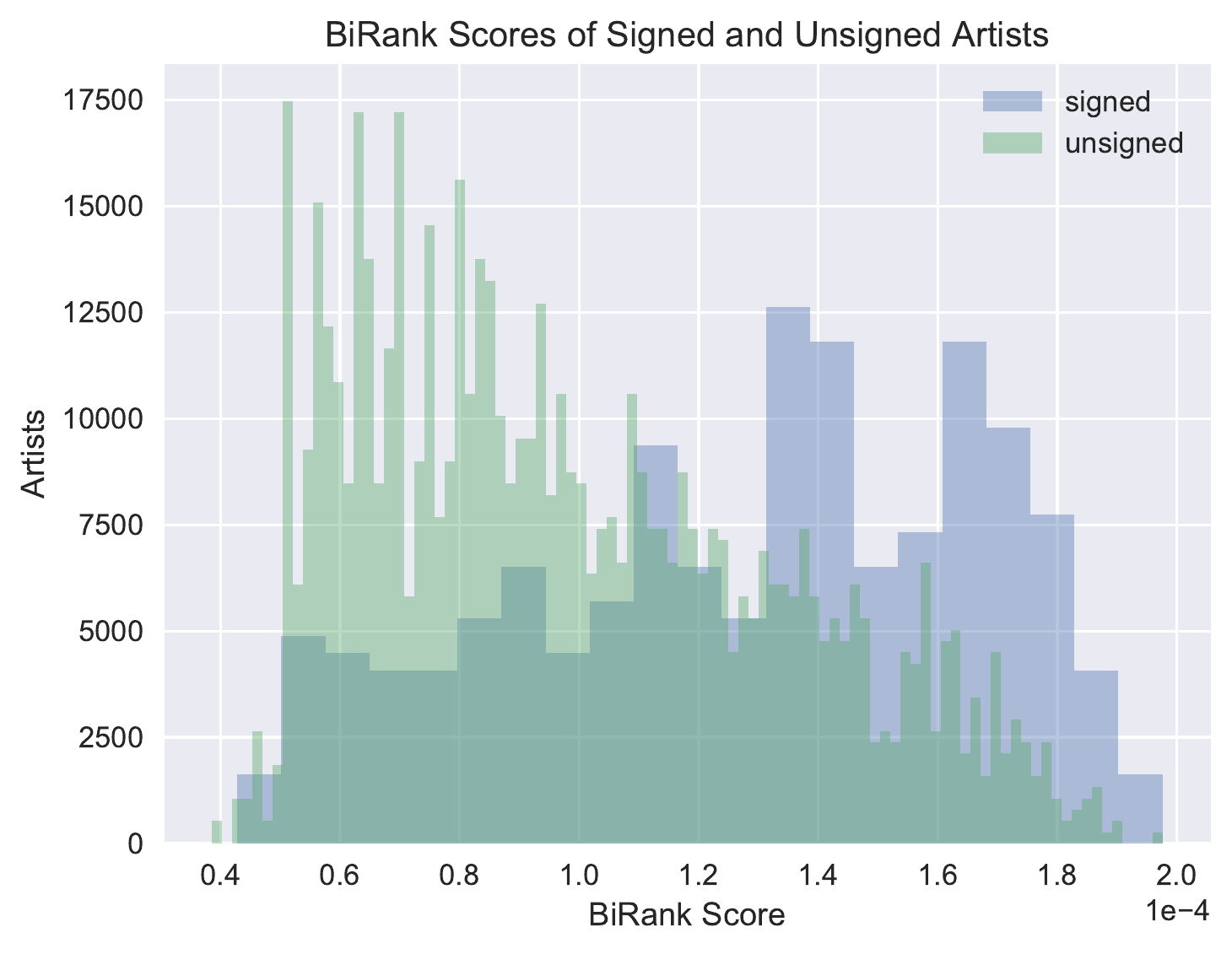}
    \caption{Histogram of signed and unsigned artists. Normalized to show relative frequency of BiRank scores. }
    \label{fig:histograms}
\end{figure}

\begin{figure}
	\includegraphics[width=0.5\textwidth]{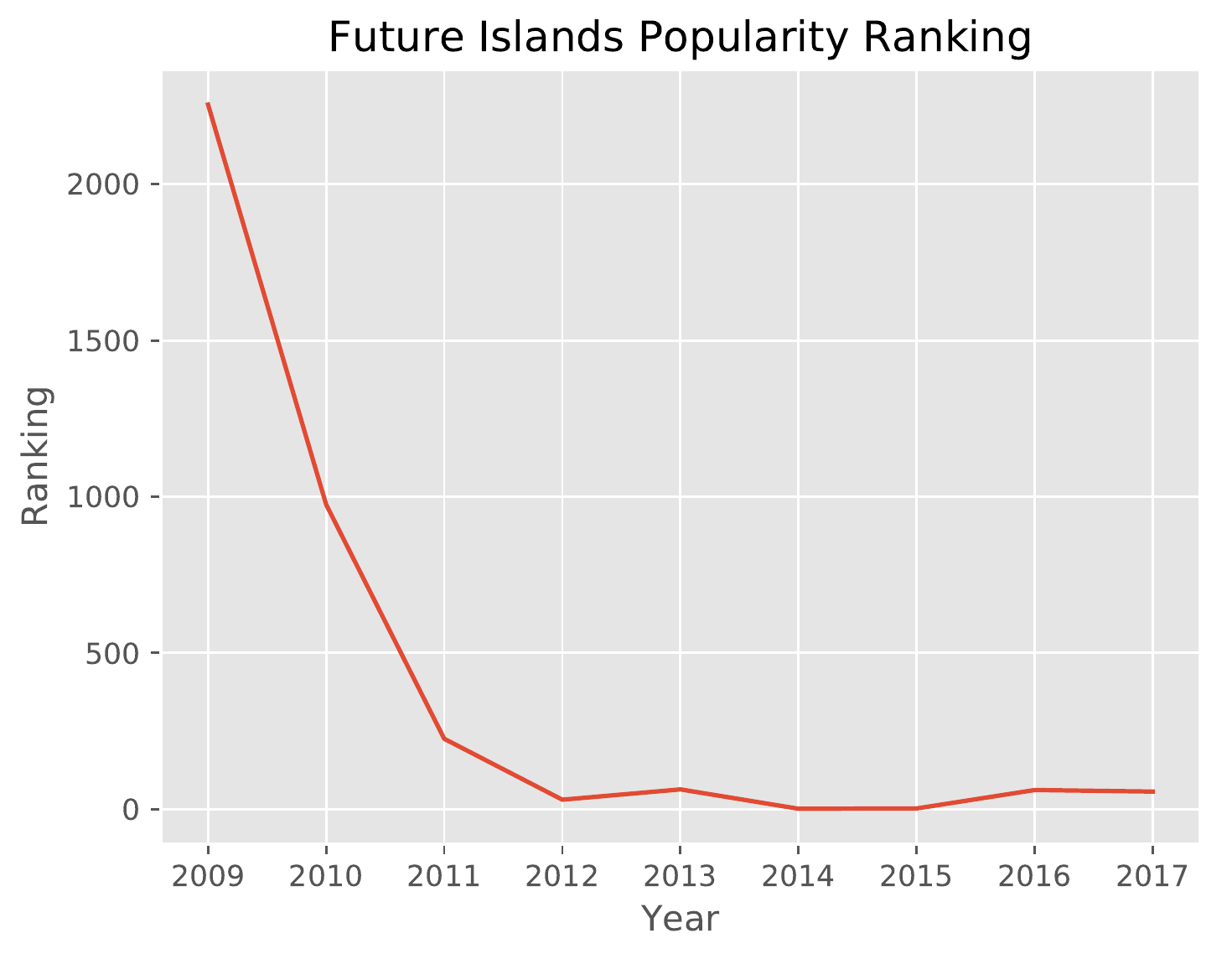}
    \caption{Trajectory of the group ``Future Islands'' through the lens of the BiRank score. The y-axis is the rank: lower is better. The BiRank score tracks the band's rise to popularity, culminating in the 2014 nomination of ``breakthrough band of the year'' by The Telegraph, suggesting that our framework can capture, and may predict, outstanding trajectories.}
    \label{fig:birankfuture}
\end{figure}

The results of this experiment can be seen in Table~\ref{tab:influencers}. These results seem to indicate promise for this method on our dataset. In the case of the venues, they correspond to some of the most popular venues in the world. As for the artists, the story is different. While they do not correspond to the most popular in terms of followers, these are the artists that have more performances in the dataset.	However, a natural question regarding the dynamics of BiRank is how indicative it is of artist success. To measure this phenomenon, we plot the histogram of BiRank scores for both signed and unsigned artists. This can be seen in Figure~\ref{fig:histograms}, where we see that the signed artists tend to have a higher BiRank score than unsigned artists.

The BiRank scores can also be useful for measuring the trajectory of an artist. By calculating the BiRank scores as previously indicated every year, with a three year moving window, we can observe the ranking of artists at different points in time. An example of this phenomenon can be seen in Figure~\ref{fig:birankfuture}. This figure shows the BiRank ranking of the artist ``Future Island'' over time. We can see that their ranking begins around the 2,300 mark. Over the course of the next years, their ranking dramatically improves, peaking with them being the top artist according to BiRank in 2014. This is corroborated by The Telegraph naming them the ``breakthrough band of the year.''\footnote{www.telegraph.co.uk/culture/music/music-festivals/10975049/Latitude-Festival-2014-Future-Islands-the-breakthrough-band-of-the-year.html}


\section{Conclusion}
\label{sec:conclusion}
In this paper we presented a novel dataset of artists and their live performances from Songkick. We complemented that data by information collected from Discogs, which contains full history of their recordings and releases. The dataset can be used for a variety of tasks which we exemplified by performing success forecasting and event prediction. 

We proposed an operational definition of {\em success}  - signing with a major label and/or their subsidiaries - and demonstrated that the event data contains useful information that can be leveraged  to forecast artists' success with better than baseline accuracy. Similarly, we observed that by utilizing the underlying structure of this data, one can also predict whether an artist will have a concert in a particular venue. The performance of simple baseline models that we carried out in all three tasks indicates that much better results can be achieved with more carefully designed methods.

Finally,  we illustrated how artist or venue influence can be measured based on analyzing a time-varying bipartite artist-venue graph. Specifically, we analyzed the evolution of the bipartite generalization of the Pagerank score, and demonstrated both qualitatively and quantitatively that its dynamics can be used to identify successful artists. 


As future work, it will be interesting to perform more fine-grained analysis of all three tasks examined here. For instance, the results presented here were averaged across different genres. It is plausible, however, that analysis will yield (subtle) differences when conditioned on the genre. Similarly, our preliminary analysis of event sequence (as opposed to bag of word representation of events) yielded some interesting geographic patterns, which warrant further and more detailed studies.  

Finally, we would like to point out two potentially important limitations of the present work. First, the definition of success used here, while operationally useful, is by no means comprehensive. Indeed, many  artists who work with independent labels, or specialize in commercially less-viable genres, can still have very successful and celebrated careers. And second, we note that  despite its demonstrated usefulness, the dataset presented here is not perfect and is likely to have some intrinsic biases, e.g., musicians might have varying incentives for joining  platforms such as \textit{Songkick} depending on the stage of their career. Identifying and potentially correcting for such biases is another important future task.   

\section*{Acknowledgements} {
This research was supported in part 
by ARO (contract no. W911NF-12-R-0012) 
and DARPA (grant no. D16AP00115). This project does not necessarily reflect the position/policy of the Government; no official endorsement should be inferred. Approved for public release; unlimited distribution.}

\newpage

\bibliographystyle{ACM-Reference-Format}
\bibliography{bibliography}


\begin{thebibliography}{42}


\ifx \showCODEN    \undefined \def \showCODEN     #1{\unskip}     \fi
\ifx \showDOI      \undefined \def \showDOI       #1{#1}\fi
\ifx \showISBNx    \undefined \def \showISBNx     #1{\unskip}     \fi
\ifx \showISBNxiii \undefined \def \showISBNxiii  #1{\unskip}     \fi
\ifx \showISSN     \undefined \def \showISSN      #1{\unskip}     \fi
\ifx \showLCCN     \undefined \def \showLCCN      #1{\unskip}     \fi
\ifx \shownote     \undefined \def \shownote      #1{#1}          \fi
\ifx \showarticletitle \undefined \def \showarticletitle #1{#1}   \fi
\ifx \showURL      \undefined \def \showURL       {\relax}        \fi
\providecommand\bibfield[2]{#2}
\providecommand\bibinfo[2]{#2}
\providecommand\natexlab[1]{#1}
\providecommand\showeprint[2][]{arXiv:#2}

\bibitem[\protect\citeauthoryear{Acar, Dunlavy, and Kolda}{Acar
  et~al\mbox{.}}{2009}]%
        {acar2009link}
\bibfield{author}{\bibinfo{person}{Evrim Acar}, \bibinfo{person}{Daniel~M
  Dunlavy}, {and} \bibinfo{person}{Tamara~G Kolda}.}
  \bibinfo{year}{2009}\natexlab{}.
\newblock \showarticletitle{Link prediction on evolving data using matrix and
  tensor factorizations}. In \bibinfo{booktitle}{\emph{Data Mining Workshops,
  2009. ICDMW'09. IEEE International Conference on}}. IEEE,
  \bibinfo{pages}{262--269}.
\newblock


\bibitem[\protect\citeauthoryear{Agreste, De~Meo, Ferrara, Piccolo, and
  Provetti}{Agreste et~al\mbox{.}}{2015}]%
        {agreste2015analysis}
\bibfield{author}{\bibinfo{person}{Santa Agreste}, \bibinfo{person}{Pasquale
  De~Meo}, \bibinfo{person}{Emilio Ferrara}, \bibinfo{person}{Sebastiano
  Piccolo}, {and} \bibinfo{person}{Alessandro Provetti}.}
  \bibinfo{year}{2015}\natexlab{}.
\newblock \showarticletitle{Analysis of a heterogeneous social network of
  humans and cultural objects}.
\newblock \bibinfo{journal}{\emph{IEEE Transactions on Systems, Man and
  Cybernetics: Systems}} \bibinfo{volume}{45}, \bibinfo{number}{4}
  (\bibinfo{year}{2015}), \bibinfo{pages}{559--570}.
\newblock


\bibitem[\protect\citeauthoryear{Bandari, Asur, and Huberman}{Bandari
  et~al\mbox{.}}{2012}]%
        {bandari2012pulse}
\bibfield{author}{\bibinfo{person}{Roja Bandari}, \bibinfo{person}{Sitaram
  Asur}, {and} \bibinfo{person}{Bernardo~A Huberman}.}
  \bibinfo{year}{2012}\natexlab{}.
\newblock \showarticletitle{The pulse of news in social media: Forecasting
  popularity.}
\newblock \bibinfo{journal}{\emph{ICWSM}}  \bibinfo{volume}{12}
  (\bibinfo{year}{2012}), \bibinfo{pages}{26--33}.
\newblock


\bibitem[\protect\citeauthoryear{Benchettara, Kanawati, and
  Rouveirol}{Benchettara et~al\mbox{.}}{2010}]%
        {benchettara2010supervised}
\bibfield{author}{\bibinfo{person}{Nesserine Benchettara},
  \bibinfo{person}{Rushed Kanawati}, {and} \bibinfo{person}{Celine Rouveirol}.}
  \bibinfo{year}{2010}\natexlab{}.
\newblock \showarticletitle{Supervised machine learning applied to link
  prediction in bipartite social networks}. In
  \bibinfo{booktitle}{\emph{Advances in Social Networks Analysis and Mining
  (ASONAM), 2010 International Conference on}}. IEEE,
  \bibinfo{pages}{326--330}.
\newblock


\bibitem[\protect\citeauthoryear{Bliss, Frank, Danforth, and Dodds}{Bliss
  et~al\mbox{.}}{2014}]%
        {bliss2014evolutionary}
\bibfield{author}{\bibinfo{person}{Catherine~A Bliss},
  \bibinfo{person}{Morgan~R Frank}, \bibinfo{person}{Christopher~M Danforth},
  {and} \bibinfo{person}{Peter~Sheridan Dodds}.}
  \bibinfo{year}{2014}\natexlab{}.
\newblock \showarticletitle{An evolutionary algorithm approach to link
  prediction in dynamic social networks}.
\newblock \bibinfo{journal}{\emph{Journal of Computational Science}}
  \bibinfo{volume}{5}, \bibinfo{number}{5} (\bibinfo{year}{2014}),
  \bibinfo{pages}{750--764}.
\newblock


\bibitem[\protect\citeauthoryear{Buza and Galambos}{Buza and Galambos}{2013}]%
        {buza2013application}
\bibfield{author}{\bibinfo{person}{Krisztian Buza} {and} \bibinfo{person}{Ilona
  Galambos}.} \bibinfo{year}{2013}\natexlab{}.
\newblock \showarticletitle{An application of link prediction in bipartite
  graphs: Personalized blog feedback prediction}. In
  \bibinfo{booktitle}{\emph{8th Japanese-Hungarian Symposium on Discrete
  Mathematics and Its Applications June}}. \bibinfo{pages}{4--7}.
\newblock


\bibitem[\protect\citeauthoryear{Cintia, Pappalardo, and Pedreschi}{Cintia
  et~al\mbox{.}}{2013}]%
        {cintia2013engine}
\bibfield{author}{\bibinfo{person}{Paolo Cintia}, \bibinfo{person}{Luca
  Pappalardo}, {and} \bibinfo{person}{Dino Pedreschi}.}
  \bibinfo{year}{2013}\natexlab{}.
\newblock \showarticletitle{" Engine Matters": A First Large Scale Data Driven
  Study on Cyclists' Performance}. In \bibinfo{booktitle}{\emph{Data Mining
  Workshops (ICDMW), 2013 IEEE 13th International Conference on}}. IEEE,
  \bibinfo{pages}{147--153}.
\newblock


\bibitem[\protect\citeauthoryear{Cintia, Rinzivillo, and Pappalardo}{Cintia
  et~al\mbox{.}}{2015}]%
        {cintia2015network}
\bibfield{author}{\bibinfo{person}{Paolo Cintia}, \bibinfo{person}{Salvatore
  Rinzivillo}, {and} \bibinfo{person}{Luca Pappalardo}.}
  \bibinfo{year}{2015}\natexlab{}.
\newblock \showarticletitle{A network-based approach to evaluate the
  performance of football teams}. In \bibinfo{booktitle}{\emph{Machine learning
  and data mining for sports analytics workshop, Porto, Portugal}}.
\newblock


\bibitem[\protect\citeauthoryear{Clauset, Larremore, and Sinatra}{Clauset
  et~al\mbox{.}}{2017}]%
        {clauset2017data}
\bibfield{author}{\bibinfo{person}{Aaron Clauset}, \bibinfo{person}{Daniel~B
  Larremore}, {and} \bibinfo{person}{Roberta Sinatra}.}
  \bibinfo{year}{2017}\natexlab{}.
\newblock \showarticletitle{Data-driven predictions in the science of science}.
\newblock \bibinfo{journal}{\emph{Science}} \bibinfo{volume}{355},
  \bibinfo{number}{6324} (\bibinfo{year}{2017}), \bibinfo{pages}{477--480}.
\newblock


\bibitem[\protect\citeauthoryear{Clauset, Moore, and Newman}{Clauset
  et~al\mbox{.}}{2008}]%
        {clauset2008hierarchical}
\bibfield{author}{\bibinfo{person}{Aaron Clauset}, \bibinfo{person}{Cristopher
  Moore}, {and} \bibinfo{person}{Mark~EJ Newman}.}
  \bibinfo{year}{2008}\natexlab{}.
\newblock \showarticletitle{Hierarchical structure and the prediction of
  missing links in networks}.
\newblock \bibinfo{journal}{\emph{Nature}} \bibinfo{volume}{453},
  \bibinfo{number}{7191} (\bibinfo{year}{2008}), \bibinfo{pages}{98--101}.
\newblock


\bibitem[\protect\citeauthoryear{Dellarocas, Zhang, and Awad}{Dellarocas
  et~al\mbox{.}}{2007}]%
        {dellarocas2007exploring}
\bibfield{author}{\bibinfo{person}{Chrysanthos Dellarocas},
  \bibinfo{person}{Xiaoquan~Michael Zhang}, {and} \bibinfo{person}{Neveen~F
  Awad}.} \bibinfo{year}{2007}\natexlab{}.
\newblock \showarticletitle{Exploring the value of online product reviews in
  forecasting sales: The case of motion pictures}.
\newblock \bibinfo{journal}{\emph{Journal of Interactive marketing}}
  \bibinfo{volume}{21}, \bibinfo{number}{4} (\bibinfo{year}{2007}),
  \bibinfo{pages}{23--45}.
\newblock


\bibitem[\protect\citeauthoryear{Evans et~al\mbox{.}}{Evans
  et~al\mbox{.}}{2013}]%
        {evans2013constitutes}
\bibfield{author}{\bibinfo{person}{M Evans} {et~al\mbox{.}}}
  \bibinfo{year}{2013}\natexlab{}.
\newblock \showarticletitle{'What Constitutes Artist Success in the Australian
  Music Industries?'}.
\newblock \bibinfo{journal}{\emph{International Journal of Music Business
  Research (IJMBR)}} (\bibinfo{year}{2013}).
\newblock


\bibitem[\protect\citeauthoryear{Ferrara, Interdonato, and Tagarelli}{Ferrara
  et~al\mbox{.}}{2014}]%
        {ferrara2014online}
\bibfield{author}{\bibinfo{person}{Emilio Ferrara}, \bibinfo{person}{Roberto
  Interdonato}, {and} \bibinfo{person}{Andrea Tagarelli}.}
  \bibinfo{year}{2014}\natexlab{}.
\newblock \showarticletitle{Online popularity and topical interests through the
  lens of instagram}. In \bibinfo{booktitle}{\emph{Proceedings of the 25th ACM
  conference on Hypertext and social media}}. ACM, \bibinfo{pages}{24--34}.
\newblock


\bibitem[\protect\citeauthoryear{Fortunato, Bergstrom, B{\"o}rner, Evans,
  Helbing, Milojevi{\'c}, Petersen, Radicchi, Sinatra, Uzzi,
  et~al\mbox{.}}{Fortunato et~al\mbox{.}}{2018}]%
        {fortunato2018science}
\bibfield{author}{\bibinfo{person}{Santo Fortunato}, \bibinfo{person}{Carl~T
  Bergstrom}, \bibinfo{person}{Katy B{\"o}rner}, \bibinfo{person}{James~A
  Evans}, \bibinfo{person}{Dirk Helbing}, \bibinfo{person}{Sta{\v{s}}a
  Milojevi{\'c}}, \bibinfo{person}{Alexander~M Petersen},
  \bibinfo{person}{Filippo Radicchi}, \bibinfo{person}{Roberta Sinatra},
  \bibinfo{person}{Brian Uzzi}, {et~al\mbox{.}}}
  \bibinfo{year}{2018}\natexlab{}.
\newblock \showarticletitle{Science of science}.
\newblock \bibinfo{journal}{\emph{Science}} \bibinfo{volume}{359},
  \bibinfo{number}{6379} (\bibinfo{year}{2018}), \bibinfo{pages}{eaao0185}.
\newblock


\bibitem[\protect\citeauthoryear{Guimer{\`a} and Sales-Pardo}{Guimer{\`a} and
  Sales-Pardo}{2009}]%
        {guimera2009missing}
\bibfield{author}{\bibinfo{person}{Roger Guimer{\`a}} {and}
  \bibinfo{person}{Marta Sales-Pardo}.} \bibinfo{year}{2009}\natexlab{}.
\newblock \showarticletitle{Missing and spurious interactions and the
  reconstruction of complex networks}.
\newblock \bibinfo{journal}{\emph{Proceedings of the National Academy of
  Sciences}} \bibinfo{volume}{106}, \bibinfo{number}{52}
  (\bibinfo{year}{2009}), \bibinfo{pages}{22073--22078}.
\newblock


\bibitem[\protect\citeauthoryear{He, Gao, Kan, and Wang}{He
  et~al\mbox{.}}{2017}]%
        {he2017birank}
\bibfield{author}{\bibinfo{person}{Xiangnan He}, \bibinfo{person}{Ming Gao},
  \bibinfo{person}{Min-Yen Kan}, {and} \bibinfo{person}{Dingxian Wang}.}
  \bibinfo{year}{2017}\natexlab{}.
\newblock \showarticletitle{Birank: Towards ranking on bipartite graphs}.
\newblock \bibinfo{journal}{\emph{IEEE Transactions on Knowledge and Data
  Engineering}} \bibinfo{volume}{29}, \bibinfo{number}{1}
  (\bibinfo{year}{2017}), \bibinfo{pages}{57--71}.
\newblock


\bibitem[\protect\citeauthoryear{Hosseinmardi, Kao, Lerman, and
  Ferrara}{Hosseinmardi et~al\mbox{.}}{2018}]%
        {hosseinmardi2018discovering}
\bibfield{author}{\bibinfo{person}{Homa Hosseinmardi},
  \bibinfo{person}{Hsien-Te Kao}, \bibinfo{person}{Kristina Lerman}, {and}
  \bibinfo{person}{Emilio Ferrara}.} \bibinfo{year}{2018}\natexlab{}.
\newblock \showarticletitle{Discovering Hidden Structure in High Dimensional
  Human Behavioral Data via Tensor Factorization}. In
  \bibinfo{booktitle}{\emph{HeteroNAM 2018: First International Workshop on
  Heterogeneous Networks Analysis and Mining}}.
\newblock


\bibitem[\protect\citeauthoryear{Judge, Higgins, Thoresen, and Barrick}{Judge
  et~al\mbox{.}}{1999}]%
        {judge1999big}
\bibfield{author}{\bibinfo{person}{Timothy~A Judge}, \bibinfo{person}{Chad~A
  Higgins}, \bibinfo{person}{Carl~J Thoresen}, {and} \bibinfo{person}{Murray~R
  Barrick}.} \bibinfo{year}{1999}\natexlab{}.
\newblock \showarticletitle{The big five personality traits, general mental
  ability, and career success across the life span}.
\newblock \bibinfo{journal}{\emph{Personnel psychology}} \bibinfo{volume}{52},
  \bibinfo{number}{3} (\bibinfo{year}{1999}), \bibinfo{pages}{621--652}.
\newblock


\bibitem[\protect\citeauthoryear{Kashima and Abe}{Kashima and Abe}{2006}]%
        {kashima2006parameterized}
\bibfield{author}{\bibinfo{person}{Hisashi Kashima} {and}
  \bibinfo{person}{Naoki Abe}.} \bibinfo{year}{2006}\natexlab{}.
\newblock \showarticletitle{A parameterized probabilistic model of network
  evolution for supervised link prediction}. In \bibinfo{booktitle}{\emph{Data
  Mining, 2006. ICDM'06. Sixth International Conference on}}. IEEE,
  \bibinfo{pages}{340--349}.
\newblock


\bibitem[\protect\citeauthoryear{Ke, Ferrara, Radicchi, and Flammini}{Ke
  et~al\mbox{.}}{2015}]%
        {ke2015defining}
\bibfield{author}{\bibinfo{person}{Qing Ke}, \bibinfo{person}{Emilio Ferrara},
  \bibinfo{person}{Filippo Radicchi}, {and} \bibinfo{person}{Alessandro
  Flammini}.} \bibinfo{year}{2015}\natexlab{}.
\newblock \showarticletitle{Defining and identifying Sleeping Beauties in
  science}.
\newblock \bibinfo{journal}{\emph{Proceedings of the National Academy of
  Sciences}} \bibinfo{volume}{112}, \bibinfo{number}{24}
  (\bibinfo{year}{2015}), \bibinfo{pages}{7426--7431}.
\newblock


\bibitem[\protect\citeauthoryear{Kennedy, Coffrin, De~Barba, and
  Corrin}{Kennedy et~al\mbox{.}}{2015}]%
        {kennedy2015predicting}
\bibfield{author}{\bibinfo{person}{Gregor Kennedy}, \bibinfo{person}{Carleton
  Coffrin}, \bibinfo{person}{Paula De~Barba}, {and} \bibinfo{person}{Linda
  Corrin}.} \bibinfo{year}{2015}\natexlab{}.
\newblock \showarticletitle{Predicting success: how learners' prior knowledge,
  skills and activities predict MOOC performance}. In
  \bibinfo{booktitle}{\emph{Proceedings of the Fifth International Conference
  on Learning Analytics And Knowledge}}. ACM, \bibinfo{pages}{136--140}.
\newblock


\bibitem[\protect\citeauthoryear{Kunegis, De~Luca, and Albayrak}{Kunegis
  et~al\mbox{.}}{2010}]%
        {kunegis2010link}
\bibfield{author}{\bibinfo{person}{J{\'e}r{\^o}me Kunegis},
  \bibinfo{person}{Ernesto~W De~Luca}, {and} \bibinfo{person}{Sahin Albayrak}.}
  \bibinfo{year}{2010}\natexlab{}.
\newblock \showarticletitle{The link prediction problem in bipartite networks}.
  In \bibinfo{booktitle}{\emph{International Conference on Information
  Processing and Management of Uncertainty in Knowledge-based Systems}}.
  Springer, \bibinfo{pages}{380--389}.
\newblock


\bibitem[\protect\citeauthoryear{Liben-Nowell and Kleinberg}{Liben-Nowell and
  Kleinberg}{2007}]%
        {liben2007link}
\bibfield{author}{\bibinfo{person}{David Liben-Nowell} {and}
  \bibinfo{person}{Jon Kleinberg}.} \bibinfo{year}{2007}\natexlab{}.
\newblock \showarticletitle{The link-prediction problem for social networks}.
\newblock \bibinfo{journal}{\emph{journal of the Association for Information
  Science and Technology}} \bibinfo{volume}{58}, \bibinfo{number}{7}
  (\bibinfo{year}{2007}), \bibinfo{pages}{1019--1031}.
\newblock


\bibitem[\protect\citeauthoryear{Ma, Sun, and Cong}{Ma et~al\mbox{.}}{2013}]%
        {ma2013predicting}
\bibfield{author}{\bibinfo{person}{Zongyang Ma}, \bibinfo{person}{Aixin Sun},
  {and} \bibinfo{person}{Gao Cong}.} \bibinfo{year}{2013}\natexlab{}.
\newblock \showarticletitle{On predicting the popularity of newly emerging
  hashtags in twitter}.
\newblock \bibinfo{journal}{\emph{Journal of the Association for Information
  Science and Technology}} \bibinfo{volume}{64}, \bibinfo{number}{7}
  (\bibinfo{year}{2013}), \bibinfo{pages}{1399--1410}.
\newblock


\bibitem[\protect\citeauthoryear{Mazloumian, Eom, Helbing, Lozano, and
  Fortunato}{Mazloumian et~al\mbox{.}}{2011}]%
        {mazloumian2011citation}
\bibfield{author}{\bibinfo{person}{Amin Mazloumian}, \bibinfo{person}{Young-Ho
  Eom}, \bibinfo{person}{Dirk Helbing}, \bibinfo{person}{Sergi Lozano}, {and}
  \bibinfo{person}{Santo Fortunato}.} \bibinfo{year}{2011}\natexlab{}.
\newblock \showarticletitle{How citation boosts promote scientific paradigm
  shifts and nobel prizes}.
\newblock \bibinfo{journal}{\emph{PloS one}} \bibinfo{volume}{6},
  \bibinfo{number}{5} (\bibinfo{year}{2011}), \bibinfo{pages}{e18975}.
\newblock


\bibitem[\protect\citeauthoryear{McLean, Oliver, and Wainwright}{McLean
  et~al\mbox{.}}{2010}]%
        {mclean2010myths}
\bibfield{author}{\bibinfo{person}{Rachel McLean}, \bibinfo{person}{Paul~G
  Oliver}, {and} \bibinfo{person}{David~W Wainwright}.}
  \bibinfo{year}{2010}\natexlab{}.
\newblock \showarticletitle{The myths of empowerment through information
  communication technologies: An exploration of the music industries and fan
  bases}.
\newblock \bibinfo{journal}{\emph{Management Decision}} \bibinfo{volume}{48},
  \bibinfo{number}{9} (\bibinfo{year}{2010}), \bibinfo{pages}{1365--1377}.
\newblock


\bibitem[\protect\citeauthoryear{Mitra and Gilbert}{Mitra and Gilbert}{2014}]%
        {mitra2014language}
\bibfield{author}{\bibinfo{person}{Tanushree Mitra} {and} \bibinfo{person}{Eric
  Gilbert}.} \bibinfo{year}{2014}\natexlab{}.
\newblock \showarticletitle{The language that gets people to give: Phrases that
  predict success on kickstarter}. In \bibinfo{booktitle}{\emph{Proceedings of
  the 17th ACM conference on Computer supported cooperative work \& social
  computing}}. ACM, \bibinfo{pages}{49--61}.
\newblock


\bibitem[\protect\citeauthoryear{Page, Brin, Motwani, and Winograd}{Page
  et~al\mbox{.}}{1999}]%
        {page1999pagerank}
\bibfield{author}{\bibinfo{person}{Lawrence Page}, \bibinfo{person}{Sergey
  Brin}, \bibinfo{person}{Rajeev Motwani}, {and} \bibinfo{person}{Terry
  Winograd}.} \bibinfo{year}{1999}\natexlab{}.
\newblock \bibinfo{booktitle}{\emph{The PageRank citation ranking: Bringing
  order to the web.}}
\newblock \bibinfo{type}{{T}echnical {R}eport}. \bibinfo{institution}{Stanford
  InfoLab}.
\newblock


\bibitem[\protect\citeauthoryear{Park, Ciampaglia, and Ferrara}{Park
  et~al\mbox{.}}{2016}]%
        {park2016style}
\bibfield{author}{\bibinfo{person}{Jaehyuk Park},
  \bibinfo{person}{Giovanni~Luca Ciampaglia}, {and} \bibinfo{person}{Emilio
  Ferrara}.} \bibinfo{year}{2016}\natexlab{}.
\newblock \showarticletitle{Style in the age of instagram: Predicting success
  within the fashion industry using social media}. In
  \bibinfo{booktitle}{\emph{Proceedings of the 19th ACM Conference on
  Computer-Supported Cooperative Work \& Social Computing}}. ACM,
  \bibinfo{pages}{64--73}.
\newblock


\bibitem[\protect\citeauthoryear{Pavlov and Ichise}{Pavlov and Ichise}{2007}]%
        {pavlov2007finding}
\bibfield{author}{\bibinfo{person}{Milen Pavlov} {and} \bibinfo{person}{Ryutaro
  Ichise}.} \bibinfo{year}{2007}\natexlab{}.
\newblock \showarticletitle{Finding experts by link prediction in co-authorship
  networks}. In \bibinfo{booktitle}{\emph{Proceedings of the 2nd International
  Conference on Finding Experts on the Web with Semantics-Volume 290}}.
  CEUR-WS. org, \bibinfo{pages}{42--55}.
\newblock


\bibitem[\protect\citeauthoryear{Pedregosa, Varoquaux, Gramfort, Michel,
  Thirion, Grisel, Blondel, Prettenhofer, Weiss, Dubourg, Vanderplas, Passos,
  Cournapeau, Brucher, Perrot, and Duchesnay}{Pedregosa et~al\mbox{.}}{2011}]%
        {scikit-learn}
\bibfield{author}{\bibinfo{person}{F. Pedregosa}, \bibinfo{person}{G.
  Varoquaux}, \bibinfo{person}{A. Gramfort}, \bibinfo{person}{V. Michel},
  \bibinfo{person}{B. Thirion}, \bibinfo{person}{O. Grisel},
  \bibinfo{person}{M. Blondel}, \bibinfo{person}{P. Prettenhofer},
  \bibinfo{person}{R. Weiss}, \bibinfo{person}{V. Dubourg}, \bibinfo{person}{J.
  Vanderplas}, \bibinfo{person}{A. Passos}, \bibinfo{person}{D. Cournapeau},
  \bibinfo{person}{M. Brucher}, \bibinfo{person}{M. Perrot}, {and}
  \bibinfo{person}{E. Duchesnay}.} \bibinfo{year}{2011}\natexlab{}.
\newblock \showarticletitle{Scikit-learn: Machine Learning in {P}ython}.
\newblock \bibinfo{journal}{\emph{Journal of Machine Learning Research}}
  \bibinfo{volume}{12} (\bibinfo{year}{2011}), \bibinfo{pages}{2825--2830}.
\newblock


\bibitem[\protect\citeauthoryear{Perozzi, Al-Rfou, and Skiena}{Perozzi
  et~al\mbox{.}}{2014}]%
        {perozzi2014deepwalk}
\bibfield{author}{\bibinfo{person}{Bryan Perozzi}, \bibinfo{person}{Rami
  Al-Rfou}, {and} \bibinfo{person}{Steven Skiena}.}
  \bibinfo{year}{2014}\natexlab{}.
\newblock \showarticletitle{Deepwalk: Online learning of social
  representations}. In \bibinfo{booktitle}{\emph{Proceedings of the 20th ACM
  SIGKDD international conference on Knowledge discovery and data mining}}.
  ACM, \bibinfo{pages}{701--710}.
\newblock


\bibitem[\protect\citeauthoryear{Rossetti, Milli, Giannotti, and
  Pedreschi}{Rossetti et~al\mbox{.}}{2017}]%
        {rossetti2017forecasting}
\bibfield{author}{\bibinfo{person}{Giulio Rossetti}, \bibinfo{person}{Letizia
  Milli}, \bibinfo{person}{Fosca Giannotti}, {and} \bibinfo{person}{Dino
  Pedreschi}.} \bibinfo{year}{2017}\natexlab{}.
\newblock \showarticletitle{Forecasting success via early adoptions analysis: A
  data-driven study}.
\newblock \bibinfo{journal}{\emph{PloS one}} \bibinfo{volume}{12},
  \bibinfo{number}{12} (\bibinfo{year}{2017}), \bibinfo{pages}{e0189096}.
\newblock


\bibitem[\protect\citeauthoryear{Sapienza, Peng, and Ferrara}{Sapienza
  et~al\mbox{.}}{2017}]%
        {sapienza2017performance}
\bibfield{author}{\bibinfo{person}{Anna Sapienza}, \bibinfo{person}{Hao Peng},
  {and} \bibinfo{person}{Emilio Ferrara}.} \bibinfo{year}{2017}\natexlab{}.
\newblock \showarticletitle{Performance Dynamics and Success in Online Games}.
  In \bibinfo{booktitle}{\emph{2017 IEEE International Conference on Data
  Mining Workshops (ICDMW)}}. \bibinfo{pages}{902--909}.
\newblock


\bibitem[\protect\citeauthoryear{Sinatra, Wang, Deville, Song, and
  Barab{\'a}si}{Sinatra et~al\mbox{.}}{2016}]%
        {sinatra2016quantifying}
\bibfield{author}{\bibinfo{person}{Roberta Sinatra}, \bibinfo{person}{Dashun
  Wang}, \bibinfo{person}{Pierre Deville}, \bibinfo{person}{Chaoming Song},
  {and} \bibinfo{person}{Albert-L{\'a}szl{\'o} Barab{\'a}si}.}
  \bibinfo{year}{2016}\natexlab{}.
\newblock \showarticletitle{Quantifying the evolution of individual scientific
  impact}.
\newblock \bibinfo{journal}{\emph{Science}} \bibinfo{volume}{354},
  \bibinfo{number}{6312} (\bibinfo{year}{2016}), \bibinfo{pages}{aaf5239}.
\newblock


\bibitem[\protect\citeauthoryear{Szabo and Huberman}{Szabo and
  Huberman}{2010}]%
        {szabo2010predicting}
\bibfield{author}{\bibinfo{person}{Gabor Szabo} {and}
  \bibinfo{person}{Bernardo~A Huberman}.} \bibinfo{year}{2010}\natexlab{}.
\newblock \showarticletitle{Predicting the popularity of online content}.
\newblock \bibinfo{journal}{\emph{Commun. ACM}} \bibinfo{volume}{53},
  \bibinfo{number}{8} (\bibinfo{year}{2010}), \bibinfo{pages}{80--88}.
\newblock


\bibitem[\protect\citeauthoryear{Ternovski and Yasseri}{Ternovski and
  Yasseri}{2017}]%
        {ternovski2017social}
\bibfield{author}{\bibinfo{person}{John Ternovski} {and} \bibinfo{person}{Taha
  Yasseri}.} \bibinfo{year}{2017}\natexlab{}.
\newblock \showarticletitle{Social Complex Contagion in Music Listenership: A
  Natural Experiment with 1.3 Million Participants}.
\newblock \bibinfo{journal}{\emph{arXiv preprint arXiv:1711.05701}}
  (\bibinfo{year}{2017}).
\newblock


\bibitem[\protect\citeauthoryear{Wang, Song, and Barab{\'a}si}{Wang
  et~al\mbox{.}}{2013}]%
        {wang2013quantifying}
\bibfield{author}{\bibinfo{person}{Dashun Wang}, \bibinfo{person}{Chaoming
  Song}, {and} \bibinfo{person}{Albert-L{\'a}szl{\'o} Barab{\'a}si}.}
  \bibinfo{year}{2013}\natexlab{}.
\newblock \showarticletitle{Quantifying long-term scientific impact}.
\newblock \bibinfo{journal}{\emph{Science}} \bibinfo{volume}{342},
  \bibinfo{number}{6154} (\bibinfo{year}{2013}), \bibinfo{pages}{127--132}.
\newblock


\bibitem[\protect\citeauthoryear{Yucesoy and Barab{\'a}si}{Yucesoy and
  Barab{\'a}si}{2016}]%
        {yucesoy2016untangling}
\bibfield{author}{\bibinfo{person}{Burcu Yucesoy} {and}
  \bibinfo{person}{Albert-L{\'a}szl{\'o} Barab{\'a}si}.}
  \bibinfo{year}{2016}\natexlab{}.
\newblock \showarticletitle{Untangling performance from success}.
\newblock \bibinfo{journal}{\emph{EPJ Data Science}} \bibinfo{volume}{5},
  \bibinfo{number}{1} (\bibinfo{year}{2016}), \bibinfo{pages}{17}.
\newblock


\bibitem[\protect\citeauthoryear{Yucesoy, Wang, Huang, and
  Barab{\'a}si}{Yucesoy et~al\mbox{.}}{2018}]%
        {yucesoy2018success}
\bibfield{author}{\bibinfo{person}{Burcu Yucesoy}, \bibinfo{person}{Xindi
  Wang}, \bibinfo{person}{Junming Huang}, {and}
  \bibinfo{person}{Albert-L{\'a}szl{\'o} Barab{\'a}si}.}
  \bibinfo{year}{2018}\natexlab{}.
\newblock \showarticletitle{Success in books: a big data approach to
  bestsellers}.
\newblock \bibinfo{journal}{\emph{EPJ Data Science}} \bibinfo{volume}{7},
  \bibinfo{number}{1} (\bibinfo{year}{2018}), \bibinfo{pages}{7}.
\newblock


\bibitem[\protect\citeauthoryear{Zhu, Guo, Yin, Ver~Steeg, and Galstyan}{Zhu
  et~al\mbox{.}}{2016}]%
        {zhu2016scalable}
\bibfield{author}{\bibinfo{person}{Linhong Zhu}, \bibinfo{person}{Dong Guo},
  \bibinfo{person}{Junming Yin}, \bibinfo{person}{Greg Ver~Steeg}, {and}
  \bibinfo{person}{Aram Galstyan}.} \bibinfo{year}{2016}\natexlab{}.
\newblock \showarticletitle{Scalable temporal latent space inference for link
  prediction in dynamic social networks}.
\newblock \bibinfo{journal}{\emph{IEEE Transactions on Knowledge and Data
  Engineering}} \bibinfo{volume}{28}, \bibinfo{number}{10}
  (\bibinfo{year}{2016}), \bibinfo{pages}{2765--2777}.
\newblock


\bibitem[\protect\citeauthoryear{Zuber, Zibung, and Conzelmann}{Zuber
  et~al\mbox{.}}{2015}]%
        {zuber2015motivational}
\bibfield{author}{\bibinfo{person}{Claudia Zuber}, \bibinfo{person}{Marc
  Zibung}, {and} \bibinfo{person}{Achim Conzelmann}.}
  \bibinfo{year}{2015}\natexlab{}.
\newblock \showarticletitle{Motivational patterns as an instrument for
  predicting success in promising young football players}.
\newblock \bibinfo{journal}{\emph{Journal of sports sciences}}
  \bibinfo{volume}{33}, \bibinfo{number}{2} (\bibinfo{year}{2015}),
  \bibinfo{pages}{160--168}.
\newblock


\end{thebibliography}

\end{document}